\documentclass[journal]{IEEEtran}
\ifCLASSINFOpdf
\else
\fi
\usepackage{amsmath}
\usepackage{graphicx}
\begin{document}
%
\title{Analysis of Dual-Hop AF Relay Systems in Mixed RF and FSO Links}
%
%
%

\author{Chao Wei and Zaichen Zhang, \emph{Senior Member, IEEE}
\thanks{The authors are with the National Mobile Communications Research Laboratory, Southeast University, Nanjing 210096, China (emails: weichao@seu.edu.cn; zczhang@seu.edu.cn).}}
\maketitle

\begin{abstract}
We analysis the performances for the dual-hop fixed gain amplify-and-forward (AF) relay systems operating over mixed radio-frequency (RF) and free-space optical (FSO) links. The RF link is subject to Rician fading and the FSO link experiences Gamma-Gamma turbulence fading. We derive the closed-form expressions for the outage probability and the average symbol error rate in terms of the Meijer's G function. All analytical results are corroborated by simulation results and the effects of fading parameters on the system performances are also studied.
\end{abstract}


%
\IEEEpeerreviewmaketitle

\section{Introduction}
%
%
%
%
\IEEEPARstart{R}{elay} scheme in recent years has become a promising solution for future wireless systems, owing to its significant cooperative diversity gain [1]. In general, the most popular and studied relaying protocols are amplify-and-forward (AF) and decode-and-forward (DF). In contrast to AF relay, DF relay requires more processing power and complexity. Extensive research has been conducted on the performances of dual-hop AF systems operating in symmetric or asymmetric links (e.g., Rayleigh links [2], Rician links [3], Nakagami-m links [4], Rayleigh and Rician links [5], and Nakagami-m and Rician links [6], etc.).

  Free-space optical (FSO) communications are line-of-sight (LOS) optical transmission techniques, offering higher bandwidth
capacity when compared to radio-frequency (RF) communications [7]. Relay scheme is also generalized to FSO communications to mitigate the degrading effects of the atmosphere turbulence-induced fading [8].

  The model proposed in [9] is an asymmetric dual-hop AF system with mixed RF/FSO links, which can employ the advantages of RF
and FSO communications simultaneously. RF links provide users with mobility and FSO links offer higher bandwidth capacity. In [9], the outage probability of a dual-hop AF system operating over the RF Rayleigh link and the FSO Gamma-Gamma turbulence link is investigated, and closed-form outage probability expression has been obtained in terms of Meijer's G function. Based on the same model proposed in [9], [10] gives new exact closed-form expressions for the cumulative distribution function, probability density function, moment generating function, and moments of the end-to-end signal-to-noise ratio in terms of the Meijer��s G function. In [11], the performance analysis of a dual-hop AF system with pointing errors of the FSO link is presented, which works in mixed Rayleigh link and Gamma-Gamma turbulence link and the closed-form expressions for outage probability, probability density function (PDF), and moment generating function (MGF) have been derived in terms of Meijer's G function.
\begin{figure}[!h]
\centering
\includegraphics[width=3.5in]{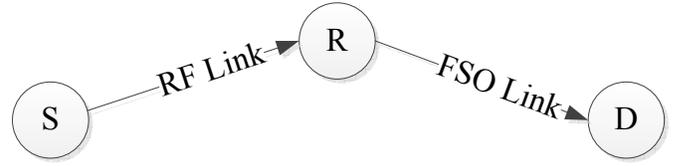}
\caption{A dual-hop AF relay system in mixed RF and FSO links.}
\label{fig_sim}
\end{figure}

  In this letter, based on the model proposed in [9], we firstly study the outage probability and the average symbol error rate
(SER) of a dual-hop fixed gain AF system with mixed Rician fading and Gamma-Gamma fading links. We derive the closed-form expressions for the PDF, outage probability (or cumulative distribution function (CDF)) and the average SER in terms of Meijer's G function.



\section{System and Channel Models}
The system under study is an asymmetric dual-hop AF relay system with no direct link, as shown in Fig.1. The source node S and the destination node D communicate through a relay node R. The first link S-R and the second link R-D are an RF link and an FSO link, respectively. We consider a fixed gain relay. Assume that node S sends a subcarrier signal to node R. At node R, the received signal can be expressed as
   \begin{equation}r_1(t)=\alpha_1s_1(t)+n_1(t)\end{equation}
where \(\alpha_1\) represents the fading gain of the RF link between nodes S and R. \(s_1(t)\) has an average power \(P_s\) , and \(n_1(t)\) is the additive white Gaussian noise (AWGN) with one-side power spectrum density (PSD) of \(N_{01}\) . At relay node R, the RF signal \(r_1(t)\) is used to modulate the irradiance of a continuous wave optical beam at the laser transmitter after being properly biased, then the retransmitted signal can be written as
   \begin{equation}s_2(t)=G(1+\xi r_1(t))\end{equation}
where \(G\) is the fixed relay gain and \(\xi\) is the modulation index satisfying the condition \(-1\leq\xi r_1(t)\leq1\) in order to avoid overmodulation. For an atmosphere turbulence channel, the received signal at destination node D after direct detection using photodetector can be written as
   \begin{equation}r_2(t)=VIAs_2(t)+n_2(t)\end{equation}
where \(V\) is the responsivity of photodetector, \(I\) is assumed to be a stationary random variable caused by atmospheric turbulence, \(A\) is the area of photodetector, and \(n_2(t)\) is the AWGN with one-side PSD of \(N_{02}\). After removing the direct current (DC) component, the instantaneous end-to-end SNR at the destination node D is given by [2]
   \begin{equation}\gamma=\frac{\gamma_1\gamma_2}{\gamma_2+C}\end{equation}
where \(\gamma_1=P_s\alpha_1^2 \slash N_{01}\) is the RF link SNR, \(\gamma_2=(VA\xi)^2I^2 \slash N_{02}\) is the FSO link SNR, and \(C=(G^2N_{01})^{-1}\).

  Assuming that the RF link is subject to Rician fading, the PDF of \(\gamma_1\) is
given by [12, Eq. 2. 16]
\begin{eqnarray}
\nonumber {f_{{\gamma _1}}}\left( r \right) = \left( {\frac{{1 + {K_1}}}{{\overline {{\gamma _1}} }}} \right)\exp \left( { - {K_1}} \right)\exp \left( { - \frac{{1 + {K_1}}}{{\overline {{\gamma _1}} }}} \right)\\
\times {I_0}\left( {2\sqrt {{K_1}\frac{{1 + {K_1}}}{{\overline {{\gamma _1}} }}r} } \right),r \ge 0
\end{eqnarray}
where \(\overline {{\gamma _1}}  = {{{P_s}} \mathord{\left/
 {\vphantom {{{P_s}} {{N_{01}}}}} \right.
 \kern-\nulldelimiterspace} {{N_{01}}}}\) is the RF link's average SNR, \(K_1\) is the ratio of the power of the LOS component to the average power of the scattered component and \(I_{\alpha}(\cdot)\) is the \(\alpha^{th}\)-order modified Bessel function of the first kind [13, Eq. 8.406.1]. The CDF of \(\gamma_1\) is given by
  \begin{equation}
  {F_{{\gamma _1}}}(r) = 1 - {Q_1}\left( {\sqrt {2{K_1}} ,\sqrt {2\frac{{1 + {K_1}}}{{\overline {{\gamma _1}} }}r} } \right),r \ge 0
  \end{equation}
where \(Q_i(\alpha,\beta)\) is the \(i^{th}\)-order Marcum Q-function [12, Eq. 4. 60]. For the special case of Rayleigh fading, the PDF and CDF of \(\gamma_1\) can be obtained by substituting \(K_1=0\) into (5) and (6), respectively. We assume that the FSO link experiences Gamma-Gamma fading, so the PDF of \(\gamma_2\) is given by [9]
  \begin{eqnarray}
  \nonumber f_{\gamma_2}(r)=\frac{(\alpha\beta)^{(\alpha+\beta)\slash2}r^{(\alpha+\beta)\slash4-1}}
  {\Gamma(\alpha)\Gamma(\beta)\overline{\gamma_2}^{(\alpha+\beta)\slash4}}\\
  \times {K_{\alpha  - \beta }}\left( {2\sqrt {\alpha \beta \sqrt {\frac{r}{{\overline {{\gamma _2}} }}} } } \right),r \ge 0
 \end{eqnarray}
 where \(\overline {{\gamma _2}}  = {{{{\left( {VA\xi } \right)}^2}} \mathord{\left/
 {\vphantom {{{{\left( {KA\xi } \right)}^2}} {{N_{02}}}}} \right.
 \kern-\nulldelimiterspace} {{N_{02}}}}\) is the FSO link��s average SNR, \(\Gamma(\cdot)\) is the Gamma function [13. Eq. 8. 310. 1] and \(K_{\alpha-\beta }(\cdot)\) is the \((\alpha-\beta)^{th}\)-order modified Bessel function of the second kind [13, Eq. 8. 432. 9]. The shaping parameters \(\alpha\) and \(\beta\) are related to the Rytov variance and the relationship \(\alpha>\beta\) always holds for FSO communication applications [14].

\section{Performance Analysis}
Firstly, we develop a general framework to obtain the PDF and CDF of the instantaneous end-to-end SNR, \(\gamma\), for the
fixed gain dual-hop relay systems. The PDF of \(\gamma\) is
\begin{equation}
{f_\gamma }\left( r \right) = \int_0^\infty  {\left( {\frac{{{\gamma _2} + C}}{{{\gamma _2}}}} \right){f_{{\gamma _1}}}\left( {\frac{{{\gamma _2} + C}}{{{\gamma _2}}}r} \right){f_{{\gamma _2}}}\left( {{\gamma _2}} \right)d{\gamma _2}}
\end{equation}
The CDF is the probability that \(\gamma\) drops below an SNR threshold \(\gamma_{th}\), or mathematically
\begin{equation}
\begin{aligned}
   P_{out}&=F_{\gamma}(\gamma_{th})=P(\gamma\leq\gamma_{th})\\
  &=\int_0^\infty  {{F_{{\gamma _1}}}\left( {\frac{{{\gamma _2} + C}}{{{\gamma _2}}}{\gamma _{th}}} \right)
   {f_{{\gamma _2}}}\left( {{\gamma _2}} \right)d{\gamma _2}}
\end{aligned}
\end{equation}
Equations (8) and (9) can be used with any \({f_{{\gamma _1}}}\left( r \right)\) and \({f_{{\gamma _2}}}\left( r \right)\) to obtain the PDF and CDF of the instantaneous end-to-end SNR for the fixed gain dual-hop relay systems.
\subsection{The PDF of \(\gamma\)}
The PDF of \(\gamma\) is obtained by substituting the expressions (5) and (7) into (8) resulting in
 \begin{equation}
 \begin{aligned}
{f_\gamma }\left( r \right)&={A_0}{B_0}\exp \left( { - \frac{{1 + {K_1}}}{{\overline {{\gamma _1}} }}r} \right)\\
&\times\int_0^\infty  {\left( {1 + \frac{C}{x}} \right){x^{\frac{{\alpha  + \beta }}{4} - 1}}\exp \left( { - \frac{{1 + {K_1}}}{{\overline {{\gamma _1}} }}\frac{{Cr}}{x}} \right)} \\
&\times {I_0}\left( {2\sqrt {{K_1}\frac{{1 + {K_1}}}{{\overline {{\gamma _1}} }}\frac{{x + C}}{x}r} } \right){K_{\alpha  - \beta }}\left( {2\sqrt {\alpha \beta \sqrt {\frac{x}{{\overline {{\gamma _2}} }}} } } \right)dx
 \end{aligned}
\end{equation}
where
\begin{equation}
{A_0} = \frac{{1 + {K_1}}}{{\overline {{\gamma _1}} \exp \left( { {K_1}} \right)}}
 \end{equation}
 \begin{equation}
 {B_0} = \frac{{{{\left( {\alpha \beta } \right)}^{\frac{{\alpha  + \beta }}{2}}}}}{{\Gamma \left( \alpha  \right)\Gamma \left( \beta  \right){{\overline {{\gamma _2}} }^{\frac{{\alpha  + \beta }}{4}}}}}
 \end{equation}
 Using the infinite series representation for the Bessel function \({I_\alpha}\left(  \cdot\right)\)[13, Eq. 8. 447. 1], and the binomial theorem \({\left( {1 + x} \right)^n} = \sum\limits_{k = 0}^n {\left( \begin{array}{l}n\\k \end{array} \right){x^k}} \), we obtain
  \begin{equation}
  \begin{split}
 &\left( {1 + \frac{C}{x}} \right){I_0}\left( {2\sqrt {{K_1}\frac{{1 + {K_1}}}{{\overline {{\gamma _1}} }}\frac{{x + C}}{x}r} } \right)\\
 &= \sum\limits_{m = 0}^\infty  {\frac{1}{{{{\left( {m!} \right)}^2}}}{{\left( {{K_1}\frac{{1 + {K_1}}}{{\overline {{\gamma _1}} }}r} \right)}^m}\sum\limits_{n = 0}^{m + 1} {\left( \begin{array}{l}
m + 1\\
n
\end{array} \right)} } {C^n}{x^{ - n}}
\end{split}
\end{equation}
Substituting (13) into (10), we obtain
\begin{equation}\begin{split}
\begin{array}{l}
{f_\gamma }\left( r \right) = {A_0}{B_0}\exp \left( { - \frac{{1 + {K_1}}}{{\overline {{\gamma _1}} }}r} \right)\\
\times \sum\limits_{m = 0}^\infty  {\frac{1}{{{{\left( {m!} \right)}^2}}}{{\left( {{K_1}\frac{{1 + {K_1}}}{{\overline {{\gamma _1}} }}r} \right)}^m}\sum\limits_{n = 0}^{m + 1} {\left( \begin{array}{l}
m + 1\\
l
\end{array} \right)} } {C^n}\\
 \times \int_0^\infty  {{x^{\frac{{\alpha  + \beta }}{4} - n - 1}}\exp \left( { - \frac{{1 + {K_1}}}{{\overline {{\gamma _1}} }}\frac{{Cr}}{x}} \right)} {K_{\alpha  - \beta }}\left( {2\sqrt {\alpha \beta \sqrt {\frac{x}{{\overline {{\gamma _2}} }}} } } \right)dx
\end{array}
\end{split}
\end{equation}
By expressing the integrands of (14) in terms of Meijer's G function [15, Eq. 5], according to [15, Eq. 11], [15, Eq. 14], and using [13, Eq. 9. 31. 2] along with [15, Eq. 21], the PDF of \(\gamma\) can be obtained as
\begin{equation}\begin{split}
\begin{array}{l}
{f_\gamma }\left( r \right) = \frac{{{A_0}{B_0}}}{{4\pi }}\exp \left( { - \frac{{1 + {K_1}}}{{\overline {{\gamma _1}} }}r} \right)\\
\times \sum\limits_{m = 0}^\infty  {\frac{1}{{{{\left( {m!} \right)}^2}}}{{\left( {{K_1}\frac{{1 + {K_1}}}{{\overline {{\gamma _1}} }}r} \right)}^m}
\sum\limits_{n = 0}^{m + 1} {\left( \begin{array}{l}
m + 1\\
l
\end{array} \right)} } {C^n}\\
 \times {\left( {\frac{{\overline {{\gamma _1}} }}{{1 + {K_1}}}\frac{1}{{Cr}}} \right)^{n - {{\left( {\alpha  + \beta } \right)} \mathord{\left/
 {\vphantom {{\left( {\alpha  + \beta } \right)} 4}} \right.
 \kern-\nulldelimiterspace} 4}}}G_{0,5}^{5,0}\left( {\frac{{{{\left( {\alpha \beta } \right)}^2}\left( {1 + {K_1}} \right)Cr}}{{16\overline {{\gamma_1}} \overline {{\gamma _2}} }}\left| \begin{array}{l}
 - \\
\Lambda 1
\end{array} \right.} \right)
\end{array}
\end{split}
\end{equation}
where
\begin{equation}
\Lambda 1 = \left\{ {\frac{{\alpha  - \beta }}{4},\frac{{\alpha  - \beta  + 2}}{4},\frac{{\beta  - \alpha }}{4},\frac{{\beta  - \alpha  + 2}}{4},n - \frac{{\alpha  + \beta }}{4}} \right\}
\end{equation}
 and \(G\left(  \cdot  \right)\) is the Meijer's G function [15, Eq. 5].
\subsection{The CDF of \(\gamma\)}
  We substitute the expressions (6) and (7) into (9) to obtain the closed-form \(F_{\gamma}(\gamma_{th})\).
We use the infinite series representation for the Marcum \(Q\)-function in the CDF of the Rician fading link [12, Eq. 4.64] and the infinite series representation for the Bessel function \(I_{\alpha}(\cdot)\) [13, Eq. 8.447.1], and obtain
\begin{equation}\begin{split}
\begin{array}{l}
{F_\gamma }\left( {{\gamma _{th}}} \right) = 1{\rm{ - }}{B_0}\exp \left( { - {K_1} - \frac{{1 + {K_1}}}{{\overline {{\gamma _1}} }}{\gamma _{th}}} \right)\\
\times \sum\limits_{i = 0}^\infty  {\sum\limits_{j = 0}^\infty  {\frac{{K_1^{i + j}}}{{j!\left( {i + j} \right)!}}{{\left( {\frac{{1 + {K_1}}}{{\overline {{\gamma _1}} }}{\gamma _{th}}} \right)}^j}} } \sum\limits_{k = 0}^j {\left( \begin{array}{l}
j\\
k
\end{array} \right){C^k}} \\
 \times \int_0^\infty  {{x^{\frac{{\alpha  + \beta }}{4} - n - 1}}\exp \left( { - \frac{{1 + {K_1}}}{{\overline {{\gamma _1}} }}\frac{{C{r_{th}}}}{x}} \right)} {K_{\alpha  - \beta }}\left( {2\sqrt {\alpha \beta \sqrt {\frac{x}{{\overline {{\gamma _2}} }}} } } \right)dx
\end{array}
\end{split}
\end{equation}
Similar to deriving (15) from (14), we can obtain the closed-form CDF of \(\gamma\) as
\begin{equation}\begin{split}
\begin{array}{l}
{F_\gamma }\left( {{\gamma _{th}}} \right) = 1 - \frac{{{B_0}}}{{4\pi }}\exp \left( { - {K_1} - \frac{{1 + {K_1}}}{{\overline {{\gamma _1}} }}{\gamma _{th}}} \right)\\
 \times \sum\limits_{i = 0}^\infty  {\sum\limits_{j = 0}^\infty  {\frac{{K_1^{i + j}}}{{j!\left( {i + j} \right)!}}{{\left( {\frac{{1 + {K_1}}}{{\overline {{\gamma _1}} }}{\gamma _{th}}} \right)}^j}} } \sum\limits_{k = 0}^j {\left( \begin{array}{l}
j\\
k
\end{array} \right){C^k}} \\
 \times {\left( {\frac{{\overline {{\gamma _1}} }}{{1 + {K_1}}}\frac{1}{{C{r_{th}}}}} \right)^{k - {{\left( {\alpha  + \beta } \right)} \mathord{\left/
 {\vphantom {{\left( {\alpha  + \beta } \right)} 4}} \right.
 \kern-\nulldelimiterspace} 4}}}G_{0,5}^{5,0}\left( {\frac{{{{\left( {\alpha \beta } \right)}^2}\left( {1 + {K_1}} \right)C{r_{th}}}}{{16\overline {{\gamma _1}} \overline {{\gamma _2}} }}\left| \begin{array}{l}
 - \\
\Lambda 2
\end{array} \right.} \right)
\end{array}
\end{split}
\end{equation}
where
\begin{equation}
\Lambda 2 = \left\{ {\frac{{\alpha  - \beta }}{4},\frac{{\alpha  - \beta  + 2}}{4},\frac{{\beta  - \alpha }}{4},\frac{{\beta  - \alpha  + 2}}{4},k - \frac{{\alpha  + \beta }}{4}} \right\}
\end{equation}
and \(B_0\) is defined as (12).
\subsection{The Average SER}
  We now derive the expression for the average SER of the system, which is applicable to modulation schemes
that have a SER expression of the form as below
\begin{equation}
{P_s} = aQ\left( {\sqrt {2b\gamma } } \right)
\end{equation}
where \(a\) and \(b\) define the modulation schemes (e.g. \(a=b=1\) for binary phase shift keying (BPSK), and  , \(a=2\), \(b = {\sin ^2}\left( {\pi /M} \right)\) for M-ary phase shift keying (M-PSK)), \(Q\left( x \right) = \int_x^\infty  {{{\exp \left( {{{ - {y^2}} \mathord{\left/
 {\vphantom {{ - {y^2}} 2}} \right.\kern-\nulldelimiterspace} 2}} \right)} \mathord{\left/
 {\vphantom {{\exp \left( {{{ - {y^2}} \mathord{\left/{\vphantom {{ - {y^2}} 2}} \right.
 \kern-\nulldelimiterspace} 2}} \right)} {\sqrt {2\pi } }}} \right.\kern-\nulldelimiterspace} {\sqrt {2\pi } }}} dy\). We give
 the average SER as
 \begin{equation}
{P_s} = \frac{a}{2}\sqrt {\frac{b}{\pi }} \int_0^\infty  {\frac{{{F_r}\left( y \right)}}{{\sqrt y }}\exp \left( { - by} \right)dy}
\end{equation}
Substituting (18) into (21) and using the identities [15, Eq. 11] and [15, Eq. 21], we can get the closed-form average SER as
\begin{equation}
\begin{array}{l}
{P_s} = \frac{a}{2} - \frac{a}{{8\pi }}\sqrt {\frac{b}{\pi }} {B_0}\exp \left( { - {K_1}} \right)\\
 \times {\sum\limits_{i = 0}^\infty  {\sum\limits_{j = 0}^\infty  {\frac{{K_1^{i + j}}}{{j!\left( {i + j} \right)!}}{{\left( {\frac{{1 + {K_1}}}{{\overline {{\gamma _1}} }}} \right)}^j}\sum\limits_{k = 0}^j {\left( \begin{array}{l}
j\\
k
\end{array} \right){C^k}\left( {\frac{{\overline {{\gamma _1}} }}{{1 + {K_1}}}\frac{1}{C}} \right)} } } ^{k - {{\left( {\alpha  + \beta } \right)} \mathord{\left/
 {\vphantom {{\left( {\alpha  + \beta } \right)} 4}} \right.
 \kern-\nulldelimiterspace} 4}}}\\
 \times {\left( {b + \frac{{1 + {K_1}}}{{\overline {{\gamma _1}} }}} \right)^{k - j - {{\left( {\alpha  + \beta  + 2} \right)} \mathord{\left/
 {\vphantom {{\left( {\alpha  + \beta  + 2} \right)} 4}} \right.
 \kern-\nulldelimiterspace} 4}}}\\
 \times G_{1,5}^{5,1}\left( {\frac{{{{\left( {\alpha \beta } \right)}^2}\left( {1 + {K_1}} \right)C}}{{16\overline {{\gamma _2}} \left( {1 + {K_1} + b\overline {{\gamma _1}} } \right)}}\left| \begin{array}{l}
\frac{1}{2} - j + k - \frac{{\alpha  + \beta }}{4}\\
\Lambda 2
\end{array} \right.} \right)
\end{array}
\end{equation}
where \({B_0}\) is defined as (12) and \(\Lambda 2\) is defined as (19).
\begin{figure}[!h]
\centering
\includegraphics[width=3.5in]{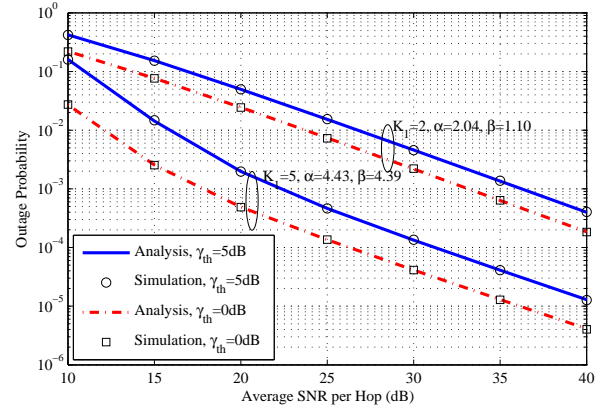}
\caption{Outage probability with threshold \(\gamma_{th} = 0,5\)dB.}
\label{fig_sim}
\end{figure}

\begin{figure}[!h]
\centering
\includegraphics[width=3.5in]{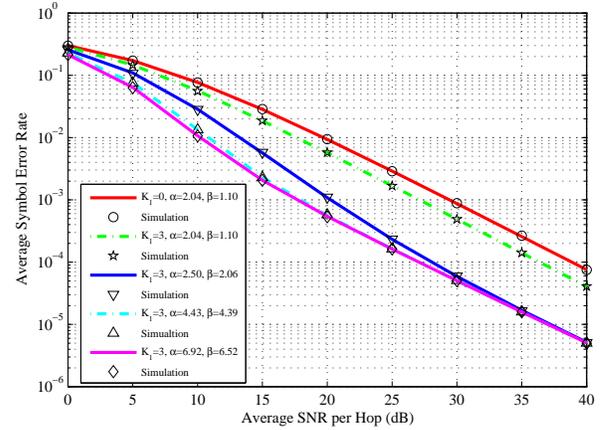}
\caption{ Average SER with fixed Rician parameter \(K_1=3\). The lines represent the analytical results.}
\label{fig_sim}
\end{figure}


\section{Numerical and Simulation Results}
  In this section, we compare the theoretical results in Section III with the simulations results. We
assume equal noise PSD, \(N_0\), at the relay and the destination, and the balanced links, \(\overline {{\gamma _1}}  = \overline {{\gamma _2}}  = \overline \gamma  \). \(C\) is set to 1. Each of the infinite summations in (18) and (22) was truncated at the \(35th\) term, because adding more terms does not affect the results in the \(15th\) decimal place.

  Fig. 2 presents the outage probability versue average SNR per hop of the dual-hop AF systems operating over two groups of
mixed fading channels. The outage threshold \(\gamma_{th}\) is set to 0 and 5dB, respectively. It is observed that the numerical results of (18) match well with the simulation results. Further, as expected, the outage probability decreases with increasing average SNR per hop and increases with severer fading channels and higher threshold values.

  Fig. 3 shows the effects of the Gamma-Gamma fading parameters, \(\alpha\) and \(\beta\), on the average SER of the dual-hop AF
systems employing BPSK modulation. Fig. 3 presents the average SER of the dual-hop AF systems with \(K_1=3\) and different \(\left( {\alpha, \beta} \right)\) pairs. \(\left( {\alpha  = 6.92, \beta  = 6.52} \right)\), \(\left( {\alpha  = 4.43, \beta  = 4.39} \right)\), \(\left( {\alpha  = 2.50, \beta  = 2.06} \right)\), and \(\left( {\alpha  = 2.04, \beta  = 1.10} \right)\) represent the weaker, weak, moderate, and strong Gamma-Gamma turbulence conditions, respectively. The average SER of a dual-hop AF system operating over Rayleigh fading link \(\left( {{K_1} = 0} \right)\) is shown for comparison. It is shown that the numerical results of (22) agree well with the simulation results. As expected, it is observed that the average SER is improved when the Gamma-Gamma turbulence condition becomes weaker. However, it is observed that the limiting slopes of the average SER curves for the different cases of turbulence conditions are the same as the limiting slope of the curve for the case of Rayleigh fading link and Gamma-Gamma parameters \(\left( {\alpha  = 2.04, \beta  = 1.10} \right)\). Moreover, Fig. 3 presents that the average SER improvement can become negligible when the turbulence conditions become weaker.

  Fig. 4 shows the effects of Rician fading parameter \(K_1\) on the average SER of the dual-hop AF systems employing BPSK
modulation. The dual-hop AF system with fixed Gamma-Gamma parameters \(\left( {\alpha  = 2.50, \beta  = 2.06} \right)\) and different Rician fading parameters, \(K_1\) = 2, 4, 6, and 8, are investigated. The average SER of a dual-hop AF system operating over Rayleigh fading link \(\left( {{K_1} = 0} \right)\) is shown for comparison. Fig. 4 shows that the numerical results of (22) match with the simulation results. It can be seen from Fig. 4 that the average SER is improved, as expected, with an increase of parameter \(K_1\). As observed in Fig. 3, the limiting slopes of the average SER curves for the different cases of \(K_1\) are the same as the limiting slope of the curve for the case of Rayleigh fading link. However, in contrast to the results shown in Fig. 3, there is a remarkable SNR gain when the Rician fading parameter \(K_1\) becomes larger. Based on Fig. 3 and Fig. 4, we conclude that the Rician fading parameter has a great impact on the average SER of the dual-hop fixed gain AF systems operating over the mixed Rician and Gamma-Gamma turbulence links.
\begin{figure}[!h]
\centering
\includegraphics[width=3.5in]{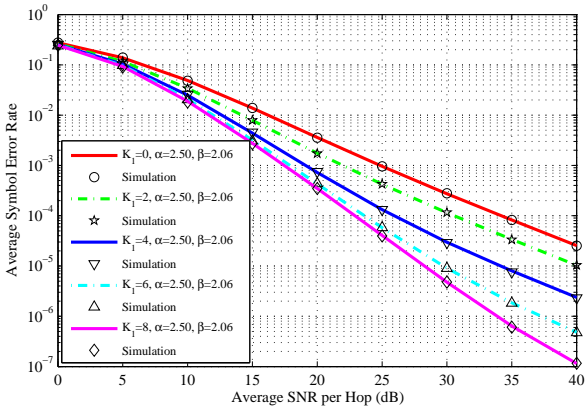}
\caption{Average SER with fixed Gamma-Gamma parameters, \(\alpha=2.50\), \(\beta=2.06\). The lines represent the analytical results.}
\label{fig_sim}
\end{figure}
\section{Conclusion}
  In this letter, we have derived the closed-form expressions for the PDF, the CDF, and the average SER of a dual-hop fixed
gain AF relaying systems operating over the mixed Rician fading and Gamma-Gamma turbulence links. The simulation results are given to validate the analytical ones. The results show that the limiting slopes of the CDF and the average SER only depend on the Rayleigh fading, irrespective of the Gamma-Gamma turbulence conditions.

\end{document}